\documentclass{Interspeech}
\usepackage{amsmath}
\usepackage{booktabs}
\usepackage{multirow}
\usepackage{graphicx}
\usepackage{xcolor}
\usepackage{pifont}
\usepackage{colortbl}

\interspeechcameraready

\title{CLEP-DG: Contrastive Learning for Speech Emotion Domain Generalization via Soft Prompt Tuning}

\author[affiliation={1}]{Jiacheng}{Shi}
\author[affiliation={1}]{Yanfu}{Zhang}
\author[affiliation={1}]{Ye}{Gao}

\affiliation{Department of Computer Science}{William \& Mary}{USA}
\email{\{jshi12,yzhang105,ygao18\}@wm.edu}

\usepackage{comment}

\begin{document}

\maketitle

\begin{abstract}
Speech Emotion Recognition (SER) is fundamental to affective computing and human-computer interaction, yet existing models struggle to generalize across diverse acoustic conditions. While Contrastive Language-Audio Pretraining (CLAP) provides strong multimodal alignment, it lacks dedicated mechanisms for capturing emotional cues, making it suboptimal for SER. To address this, we propose CLEP-DG, a framework that enhances CLAP’s robustness in emotion recognition. First, we fine-tune CLAP to obtain CLEP, adapting it on large-scale emotional speech datasets to better encode emotion-relevant features. Then, we introduce Acoustic Context Prompt Tuning (ACPT), a text-driven augmentation strategy that optimizes learnable prompt vectors to model diverse acoustic environments without additional labeled audio. Finally, leveraging cross-modal transferability, we train a classifier on text-derived embeddings and apply it to the audio encoder during inference, mitigating domain shifts between textual supervision and audio-based emotion recognition. Experiments across five benchmark datasets show that CLEP-DG outperforms prior CLAP-based approaches, achieving state-of-the-art performance in both supervised and domain generalization settings.
\end{abstract}

\section{Introduction}
\label{sec:intro}
Speech Emotion Recognition (SER) classifies emotional states from speech and acoustic expressions, serving as a cornerstone for affective computing~\cite{tao2005affective} and human-computer interaction~\cite{preece1994human}. Recent advances in Contrastive Language-Audio Pretraining (CLAP)~\cite{elizalde2023clap,wu2023large} have demonstrated strong generalization by aligning text and audio in a shared embedding space. However, existing CLAP-based models are primarily designed for speech retrieval and transcription, lacking dedicated mechanisms to capture emotional attributes. As a result, they struggle to differentiate emotional cues from variations in recording conditions and background noise, limiting effectiveness in SER and ability to model nuanced affective expressions.


Recent works have adapted CLAP for speech emotion recognition by incorporating task-specific modifications. ParaCLAP~\cite{jing2024paraclap} enhances zero-shot recognition by introducing query-based contrastive learning for improved audio-text alignment. GEmo-CLAP~\cite{pan2023gemo} integrates affective embeddings to refine multimodal representations of emotional attributes. HuBERT-CLAP~\cite{hu2024cross} leverages self-supervised speech representations to improve robustness and strengthen audio-language alignment. Despite these advancements, these methods rely on predefined textual representations and require extensive retraining with annotated emotion audio to adapt to new domains, limiting scalability and real-world applicability. A promising alternative is soft-prompt tuning~\cite{zhou2022cpl, zhou2022ltp}, which has been widely explored in vision-language models. Instead of relying on manually crafted text templates (e.g., “this is a sound of”), soft prompts introduce learnable tokens while keeping the CLAP text encoder unchanged, enabling efficient adaptation with minimal labeled data and computational overhead. Inspired by this, we investigate whether CLAP’s latent space can inherently capture diverse acoustic contexts without explicit supervision. Given that large-scale pre-trained models inherently encode varied acoustic information, could they serve as implicit proxies for multiple source domains? This leads to our central research question: \textit{Can CLAP’s generalization be further enhanced by simulating diverse acoustic contexts in its latent space while preserving its ability to capture emotional attributes?}

In this paper, we propose that diverse textual representations of acoustic contexts can effectively model variability in audio data. Our observations indicate that features extracted by the audio encoder closely align with text embeddings from corresponding descriptions, making textual features a viable training alternative. To this end, \textbf{we make three key contributions:} (1) We introduce CLEP-DG, an emotion-aware multimodal model obtained by fine-tuning CLAP on large-scale emotional speech datasets, enhancing the alignment between emotional audio and textual representations. (2) We propose Acoustic Context Prompt Tuning (ACPT), a text-driven augmentation strategy that optimizes learnable prompt vectors to encode diverse soundscape variations, improving robustness across acoustic conditions without requiring additional labeled speech data. (3) CLEP-DG achieves state-of-the-art performance on both supervised and domain generalization settings across multiple benchmark datasets, demonstrating superior adaptability to real-world acoustic variability. Additionally, leveraging cross-modal transferability~\cite{zhang2023diagnosing, dunlap2023using,cho2023promptstyler, tang2025dpstyler}, we train a classifier on text-derived features and apply it to audio embeddings during inference, effectively bridging the gap between textual supervision and audio-based emotion recognition.

\section{Related Work}
\textbf{Domain Generalization in SER.} Large-scale pretrained models like Whisper~\cite{C10}, WavLM~\cite{chen2022wavlm}, and HuBERT~\cite{hsu2021hubert} enhance SER robustness by leveraging extensive speech data. While effective, they rely on supervised learning and lack explicit modeling of acoustic variability, limiting adaptability to unseen domains. \textbf{Audio-Language Pretraining in SER.} 
Recent advancements in audio-language pretraining have enabled models to perform zero-shot classification and retrieval tasks. Wav2CLIP \cite{C14} and AudioCLIP \cite{C15} extend contrastive learning to the audio domain, yet their reliance on general acoustic features hinders their ability to capture fine-grained emotional expressions. CLAP (Laion-AI) \cite{wu2023large} and CLAP (Microsoft) \cite{elizalde2023clap} leverage large-scale datasets for multimodal learning but remain primarily optimized for speech content rather than affective understanding. CompA-CLAP \cite{C17} introduces compositional prompts to improve generalization, yet its focus on semantic variations overlooks emotion-specific adaptations. 
Recent works~\cite{jing2024paraclap,pan2023gemo,hu2024cross} have adapted CLAP for SER, yet key challenges persist. Existing approaches rely on predefined textual templates, scarce labeled emotional datasets, or supervised adaptation, limiting their flexibility, scalability, and generalization to unseen domains.
Our approach extends CLAP for SER by fine-tuning on large-scale emotional audio datasets and integrating text-driven augmentation. We employ prompt tuning to remove reliance on predefined textual representations and audio supervision, enabling more robust emotion recognition across varied acoustic environments.

\section{Proposed Method}

\begin{figure}[!t]
  \centering
  \includegraphics[scale=0.38]{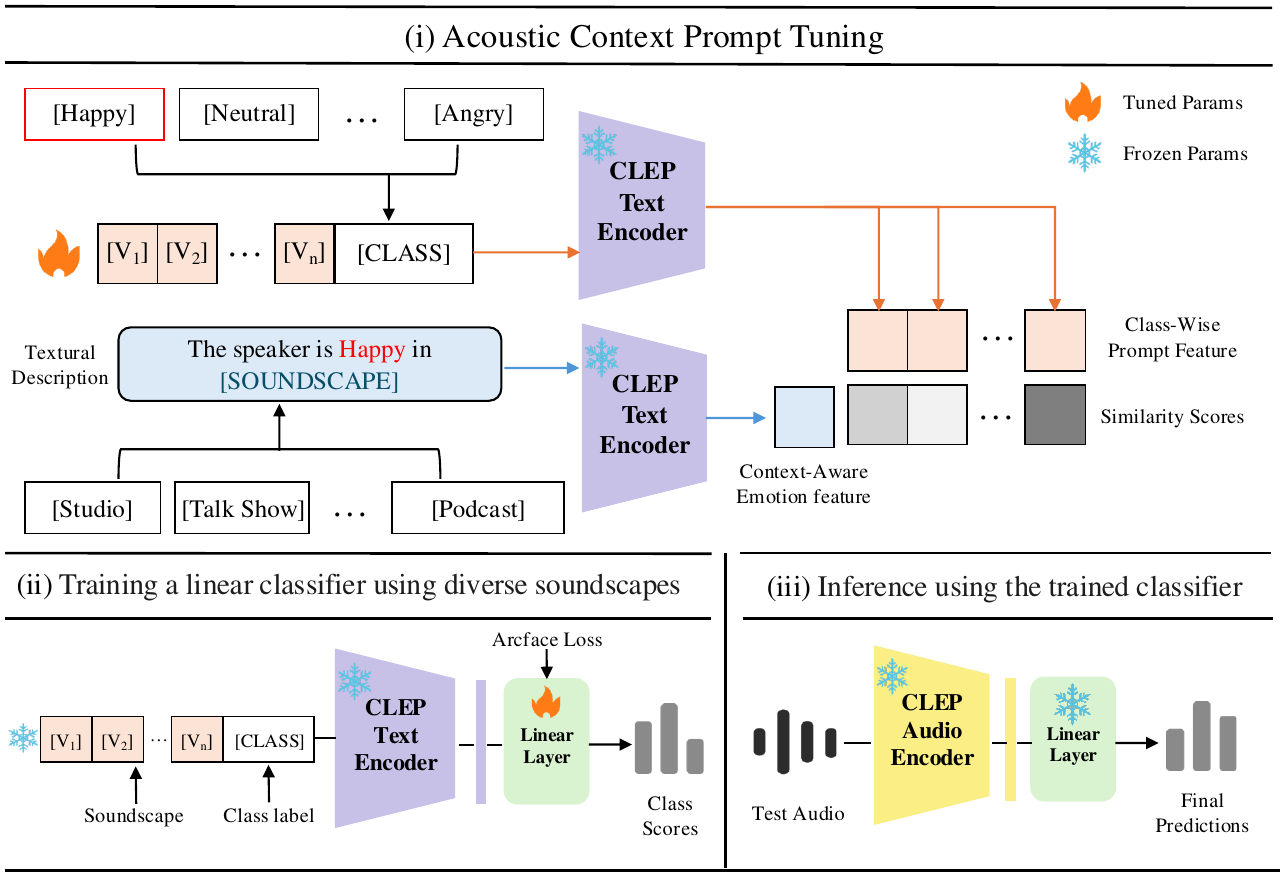}
  \caption{Illustration of the soft-prompt tuning, training a classifier and inference stage of our method.
  }
  \label{fig:pipline}
  \vspace{-4mm}
\end{figure}

\subsection{Preliminary}
The zero-shot classification in CLAP is performed by computing the similarity between an audio embedding and text-based class embeddings derived from a prompt template. Given an audio input $x_i$, its clip-level embedding $f(x_i) \in \mathbb{R}^D$ is extracted using the CLAP audio encoder $f$, where $D$ represents the joint embedding space. The class-wise text embeddings $\{w_c\}_{c=1}^{C}$ are generated by feeding prompt templates of the form  "this is a sound of [CLASS]"  into the CLAP text encoder $g$, where $C$ is the number of classes in the downstream dataset, and [CLASS] is instantiated with emotion labels such as  "Happy"  or  "Neutral" . The similarity score between an audio sample $x_i$ and a class $c$ is computed as: $\operatorname{sim}(f(x_i), w_c) = \frac{f(x_i) \cdot w_c}{\Vert f(x_i) \Vert \Vert w_c \Vert}$ where $\Vert \cdot \Vert$ denotes the $L_2$ norm. The predicted class label $\hat{y}_i$ is then obtained via: $\hat{y}_i = \arg\max_{c} \operatorname{sim}(f(x_i), w_c)$.

\subsection{Overview}
CLAP models excel in speech transcription and retrieval but lack mechanisms to capture emotional attributes, limiting their effectiveness in speech emotion recognition. To address this, we fine-tune CLAP on large-scale emotional speech datasets, yielding CLEP, an emotion-aware audio-language model. However, emotion perception is highly influenced by soundscape variations, where identical emotional cues may be interpreted differently in studio recordings or TV dialogues. Collecting diverse soundscape-labeled emotional speech for generalization remains impractical. We propose Acoustic Context Prompt Tuning (ACPT), a text-driven augmentation method that enhances CLEP’s soundscape robustness without additional audio data. ACPT generates text prompts simulating diverse soundscapes, augmenting CLEP's training distribution to improve robustness against soundscape variations. Thanks to cross-modal transferability 
phenomenon~\cite{zhang2023diagnosing, dunlap2023using,cho2023promptstyler, tang2025dpstyler}, we train a linear classifier on text-derived acoustic features extracted by a pre-trained text encoder \(T(\cdot)\). During inference, CLEP’s audio encoder \(A(\cdot)\) extracts speech features, which are then classified by the trained model. The classifier learns emotion representations from text embeddings and effectively generalizes to audio embeddings, improving adaptability across soundscapes.

\subsection{Finetuning CLAP on Emotional Audio Understanding}
We select CLAP for its strong performance across diverse speech tasks, suggesting that, with cross-modal transferability phenomenon, it can effectively extend to SER. However, CLAP, originally designed for speech transcription and retrieval, underperforms in SER due to its focus on linguistic content over affective cues. To enhance its emotion modeling, we fine-tune CLAP on large-scale emotional audio datasets as shown in Fig.~\ref{tab:datasets}.  The training data consists of emotion-labeled audio-text pairs \((X_i^a, X_i^t)\), where \(X_i^a\) is an audio sample and \(X_i^t\) is its corresponding text description. CLAP’s audio and text encoders extract representations as \(A_i = \operatorname{MLP}_a(f_a(X_i^a))\) and \(T_i = \operatorname{MLP}_t(f_t(X_i^t))\), where \(f_a(\cdot)\) and \(f_t(\cdot)\) are the audio and text encoders, and \(\operatorname{MLP}_a\), \(\operatorname{MLP}_t\) project embeddings into a shared \(D\)-dimensional space. The model is trained using a contrastive objective to align audio-text pairs: \resizebox{\linewidth}{!}{$L = \frac{1}{2N} \sum_{i=1}^{N} \left( \log \frac{\exp(A_i \cdot T_i / \tau)}{\sum_{j=1}^{N} \exp(A_i \cdot T_j / \tau)} + \log \frac{\exp(T_i \cdot A_i / \tau)}{\sum_{j=1}^{N} \exp(T_i \cdot A_j / \tau)} \right)$} where \(\tau\) is a learnable temperature parameter. Training follows the settings in~\cite{wu2023large}.  Fine-tuning on emotional audio improves CLAP’s ability to model emotion-relevant cues and strengthens audio-text alignment in the embedding space. We follow the fine-tuning setup in~\cite{wu2023large}, training only the audio encoder while freezing the text encoder to preserve its broad language understanding from large-scale pretraining.

\subsection{Acoustic Context Prompt Tuning for Soundscape}
To enhance the generalization of emotion recognition models across diverse acoustic environments, we propose Acoustic Context Prompt Tuning (ACPT), a text-driven augmentation approach that simulates variations in background conditions without requiring additional audio data. Instead of collecting and annotating emotion-labeled speech across different environments, ACPT leverages learnable prompt embeddings to enrich the text representation space, enabling the model to disentangle emotional cues from domain-specific artifacts such as background noise and recording conditions. Specifically, a prompt template is initialized as \( V = [v_1, v_2, \dots, v_N, \text{[CLASS]}] \), where \( v_j \) denotes the \( j \)-th learnable token, and \text{[CLASS]} represents an emotion category (e.g., ``Happy'' or ``Sad''). These prompts are passed through the CLEP text encoder \( T(\cdot) \), producing class-wise prompt embeddings \( u_c = T(V_c) \), where \( u_c \in \mathbb{R}^D \) and \( c \in \{1, \dots, C\} \), with \( C \) being the number of emotion categories. Given a set of textual descriptions \( \{t_m, y_m\}_{m=1}^{M} \), where \( t_m \) is a text sample describing an emotional expression and \( y_m \) is its corresponding one-hot emotion label, we optimize the classification objective as:
\begin{equation}
\mathcal{L}_{\text{CLS}} = - \sum_{m=1}^{M} \sum_{l=1}^{C} y_{m,l} \log \frac{\operatorname{sim}(t_m, u_l)/\omega}{\sum_{c=1}^{C} \operatorname{sim}(t_m, u_c)/\omega}
\end{equation} where \( \operatorname{sim}(\cdot, \cdot) \) denotes the cosine similarity function, and \( \omega \) is a temperature scaling parameter. To further enhance emotion category separation, we draw inspiration from a ranking-based contrastive loss~\cite{gong2013deep}. For each prompt \( t_m \), we define the set of positive emotion labels as \( y_m^+ = \{l | y_{m,l} = 1\} \) and negative labels as \( y_m^- = \{l | y_{m,l} = 0\} \). The ranking loss is formulated as:
\begin{equation}
\mathcal{L}_{\text{RANK}} = \sum_{m=1}^{M} \sum_{i \in y_m^+} \sum_{j \in y_m^-} \max(0, 1 - \operatorname{sim}(t_m, u_i) + \operatorname{sim}(t_m, u_j))
\end{equation} which encourages the model to maximize similarity between prompts and their corresponding emotion categories while separating non-matching ones. Finally, the overall training loss is defined as $\mathcal{L}_{\text{ACPT}} = \mathcal{L}_{\text{CLS}}+ \mathcal{L}_{\text{RANK}}$ . By optimizing both objectives, ACPT refines the alignment between textual prompts and emotion representations, improving the robustness of emotion recognition across diverse acoustic conditions.
\subsection{Cross-Modal Classifier Training and Inference}
After obtaining soundscape-conditioned text embeddings from Acoustic Context Prompt Tuning, we train a cross-modal classifier to bridge textual and audio representations. Specifically, we generate enriched embeddings by combining learned prompt representations with predefined emotion categories using the CLEP text encoder \( T(\cdot) \). These serve as training features for a linear classifier, refining its decision boundary for emotion recognition. To enhance class separability, we adopt ArcFace loss~\cite{deng2019arcface}, which introduces an additive angular margin, ensuring intra-class compactness while increasing inter-class distance, strengthening generalization across acoustic conditions.During inference, we replace text-derived embeddings with CLEP’s audio embeddings extracted via the audio encoder \( A(\cdot) \in \mathbb{R}^C \). Given an input audio sample, its representation is projected into the joint embedding space and classified by the trained cross-modal model. This process leverages cross-modal transferability~\cite{zhang2023diagnosing}, allowing the classifier to apply knowledge from text-derived representations to audio-based emotion recognition.

\section{Experiments}

\begin{table}[t]
  \centering
  \caption{The datasets for finetuning CLAP model. }
  \label{tab:datasets}
  \vspace{-2mm}
  \resizebox{\columnwidth}{!}{ 
  \begin{tabular}{l c c c c c c}
    \hline
    \textbf{Dataset} & \textbf{Emotion} & \textbf{Language} & \textbf{Speaker} & \textbf{Source} & \textbf{\#Utterances} & \textbf{\#Hours} \\
    \hline
    IEMOCAP~\cite{d1}     & 5  & English & 10   & Act        & 5531   & 7.0  \\
    MELD~\cite{d2}        & 7  & English & 407  & Friends TV & 13847  & 12.2 \\
    MEAD~\cite{d3}        & 8  & English & 60   & Act        & 31792  & 37.3 \\
    CMU-MOSEI~\cite{d4}   & 7  & English & 1000 & YouTube    & 44977  & 91.9 \\
    \hline
    \textbf{Total}  &  &  &    &          & 96147  & 148.4 \\
    \hline
  \end{tabular}
  }
  \vspace{-5mm}
\end{table}

\begin{table*}[t]
\centering
\setlength\tabcolsep{2pt}
\fontsize{7}{7}\selectfont
\resizebox{\textwidth}{!}{%
\begin{tabular}{@{}lcccccccccccc@{}}
\toprule \midrule
\multirow{2}{*}{\textbf{Model}} 
    & \multicolumn{2}{c}{\textbf{Wav2CLIP}~\cite{wu2021wav2clip}}
    & \multicolumn{2}{c}{\textbf{AudioCLIP}~\cite{guzhov2021audioclip}}
    & \multicolumn{2}{c}{\textbf{CLAP (Microsoft)}~\cite{elizalde2023clap}} 
    & \multicolumn{2}{c}{\textbf{CLAP (Laion-AI)}~\cite{wu2023large}}
    & \multicolumn{2}{c}{\textbf{CompA-CLAP}~\cite{ghosh2023compa}} 
    & \multicolumn{2}{c}{\textbf{CLEP-DG (Ours)}} \\
\cmidrule(lr){2-3}
\cmidrule(lr){4-5}
\cmidrule(lr){6-7}
\cmidrule(lr){8-9}
\cmidrule(lr){10-11}
\cmidrule(lr){12-13}
    & WA (\%) & UA (\%)  
    & WA (\%) & UA (\%)  
    & WA (\%) & UA (\%)   
    & WA (\%) & UA (\%)  
    & WA (\%) & UA (\%)  
    & WA (\%) & UA (\%)   \\
\midrule
\multicolumn{12}{l}{\textit{Supervised Learning (In-Domain Evaluation). Evaluates CLEP's classification ability.}} \\
IEMOCAP~\cite{d1}     & 58.31  & 53.16  
                       & 61.89  & 59.52  
                       & 72.54  & 70.53  
                       & 72.91  & 71.45  
                       & 73.56  & 72.14  
                       & \cellcolor[HTML]{f1e1f5} \textbf{75.18}  & \cellcolor[HTML]{f1e1f5} \textbf{74.43}  \\
MELD~\cite{d2}        & 33.57  & 15.23  
                       & 36.83  & 16.92  
                       & 47.63  & 19.61  
                       & 48.22  & 20.47  
                       & 49.08  & 20.83  
                       & \cellcolor[HTML]{f1e1f5}\textbf{50.06}  & \cellcolor[HTML]{f1e1f5}\textbf{23.28}  \\ 
\midrule \midrule
\multicolumn{12}{l}{\textit{Domain Generalization (DG). These datasets are unseen during training, evaluating the model's generalizability.}} \\
RAVDESS~\cite{d5}     & 23.91  & 19.14  
                       & 25.26  & 19.27  
                       & 35.73  & 32.85  
                       & 35.44  & 32.26  
                       & 36.52  & 34.13  
                       & \cellcolor[HTML]{f1e1f5}\textbf{39.17}  & \cellcolor[HTML]{f1e1f5}\textbf{36.58}  \\
TESS~\cite{dupuis2010toronto}  
                      & 36.82  & 33.07  
                      & 39.53  & 34.91  
                      & 49.76  & 45.93  
                      & 51.98  & 48.35  
                      & 52.37  & 49.15  
                      & \cellcolor[HTML]{f1e1f5}\textbf{54.59}  & \cellcolor[HTML]{f1e1f5} \textbf{50.71}  \\
SAVEE~\cite{d6}       & 26.54  & 20.48  
                      & 30.17  & 22.84  
                      & 39.91  & 31.63  
                      & 40.16  & 31.87  
                      & 40.36  & 32.41  
                      & \cellcolor[HTML]{f1e1f5}\textbf{43.97}  & \cellcolor[HTML]{f1e1f5}\textbf{34.53}  \\ 
\bottomrule
\end{tabular}
} 
\caption{\small Performance comparison on mainstream emotion datasets. WA(\%) and UA(\%) are averaged over three random seeds. CLEP-DG outperforms models in both supervised and domain generalization (DG) settings.}
\label{tab:main}
\vspace{-3mm}
\end{table*}

\subsection{Datasets and Model Architecture.}
\textbf{Datasets.} To adapt CLAP for emotion recognition, we fine-tune it on multiple large-scale emotional speech datasets, as shown in Tab.~\ref{tab:datasets}. Specifically, we utilize IEMOCAP~\cite{d1}, MELD~\cite{d2}, MEAD~\cite{d3}, and CMU-MOSEI~\cite{d4}, covering a diverse range of speakers, recording environments, and conversational contexts. IEMOCAP consists of dyadic conversations with scripted and improvised emotional speech from 10 speakers and is evaluated using leave-one-session-out 5-fold CV. MELD, derived from the \textit{Friends} TV series, provides multi-party conversational data with seven emotion labels and is evaluated under its original split setup. MEAD is a large-scale audio-visual dataset featuring 60 actors performing scripted monologues in eight emotional states at varying intensity levels. CMU-MOSEI is a multimodal dataset that captures sentiment and emotion expressions in online videos. To ensure consistency with prior studies~\cite{d5,d6}, we merge the ``excited'' and ``happy'' categories in IEMOCAP, resulting in a four-class classification task to match the experimental setup of~\cite{chen2023dst} and~\cite{ye2023temporal}. The RAVDESS\cite{livingstone2018ryerson}, TESS~\cite{dupuis2010toronto}, and SAVEE~\cite{haq2009speaker} datasets are evaluated under a random leave-one-out 10-fold CV setup and excluded from the fine-tuning phase to assess the model’s generalizability.

\noindent\textbf{Architecture.} We adopt CLAP, an advanced
 language-audio model published by LAION-AI~\cite{wu2023large}, which uses PANN-14~\cite{kong2020panns} as the audio encoder and RoBERTa~\cite{liu2019roberta} as the text encoder. PANN-14, a CNN-based model, employs seven downsampling and seven upsampling convolutional layers to extract hierarchical acoustic features. RoBERTa, a transformer-based language model, encodes textual inputs into dense representations. For the text encoder, we insert learnable prompt tokens after the CLS token, enabling task-specific adaptation via prompt tuning.

\subsection{Implementation Details}
\vspace{-1mm}
\textbf{Fine-tuning CLAP.} 
We utilize raw audio waveforms resampled to 16 kHz for both training and evaluation. To standardize input length, all audio sequences are truncated or zero-padded to 5 seconds. Each audio file is paired with a text label in the format “This is a [EMOTION] sound”, where \texttt{[EMOTION]} corresponds to the ground-truth emotion label (e.g., “angry,” “happy”), ensuring semantically consistent alignment between modalities. During fine-tuning, the audio encoder remains trainable, while the text encoder is frozen to preserve its broad language understanding from large-scale pretraining. This prevents overfitting to the limited text distribution of emotion datasets while ensuring robust text-audio alignment. Training is conducted using the Adam optimizer with a batch size of 64 for 80 epochs. The audio encoder is optimized with a learning rate of $1 \times 10^{-5}$, while projection layers and other trainable components use $1 \times 10^{-3}$ for stable, efficient convergence.


\noindent\textbf{Learning Soundscape-Conditioned Prompt Vectors.} 
Following prior prompt learning methodologies~\cite{zhou2022cpl, zhou2022ltp}, we construct learnable soundscape-conditioned prompts to enhance contextual diversity. Specifically, 
We select 12 soundscapes aligned with conditions shared by CLAP’s pretraining and our fine-tuning data 
and initialize their corresponding embeddings using a zero-mean Gaussian distribution with a standard deviation of 0.02. Each prompt consists of \( N_p = 8 \) learnable tokens, which are prepended to class label tokens in textual descriptions. These prompts are optimized with the $L_{ACPT}$ loss over 120 iterations. The training process employs \textit{SGD} with a learning rate of $2 \times 10^{-3}$ and a momentum coefficient of 0.9.

\noindent\textbf{Training a Linear Classifier and Inference.} 
The linear classifier is trained for 50 epochs using the SGD optimizer with a learning rate of $2 \times 10^{-3}$, momentum of 0.9, and a batch size of 16, following the training configuration in~\cite{dunlap2023using}. To enhance class separability, we employ ArcFace loss, which encourages intra-class compactness while enforcing inter-class dispersion in the feature space. During inference, audio inputs are pre-processed identically to the fine-tuning phase, including resampling to 16 kHz and adjusting to a fixed 5-second duration through truncation or padding. The CLEP-DG audio encoder $A(\cdot)$ extracts $\ell_2$-normalized embeddings, which are then passed through the trained classifier to generate final predictions.

\subsection{Performance Comparison with Existing Approaches}
\vspace{-1mm}
Since prior CLAP-based models are primarily designed for speech transcription and retrieval, we ensure a fair comparison by fine-tuning all models on the same emotional speech datasets before evaluation, as shown in Tab.~\ref{tab:main}. The higher WA compared to UA reflects class imbalance in emotion datasets. CLEP-DG outperforms the strongest baseline, CompA-CLAP, by 1.62\% and 0.98\% WA on IEMOCAP and MELD, respectively. As ParaCLAP, Emo-CLAP, and HuBERT-CLAP have not released code, direct comparisons are infeasible. All reported results are averaged over three runs with different random seeds. To further assess generalization, we evaluate CLEP-DG on three emotion datasets unseen during fine-tuning. On RAVDESS, TESS, and SAVEE, CLEP-DG improves WA by 2.65\%, 2.22\%, and 3.61\%, respectively. These results highlight the effectiveness of our text-driven augmentation in enhancing generalization by enriching the text-audio joint space with diverse acoustic contexts. By simulating variations in recording environments through textual prompts, CLEP-DG improves robustness to unseen conditions.


\subsection{Ablation Studies}
\vspace{-1mm}
\begin{table}[htbp]
  \centering
  \caption{Ablation study on the effects of fine-tuning CLAP and using the $L_{\text{ACPT}}$ loss.}
  \vspace{-3mm}
  \label{tab:ablation_finetune}
  \resizebox{\columnwidth}{!}{
  \begin{tabular}{cccccccc}
    \toprule
    Fine-tune & 
    ACPT &
    IEMOCAP & 
    MELD & 
    RAVDESS & 
    TESS & 
    SAVEE & 
    Avg. \\
    \midrule
    $-$ & $-$ & 36.35 & 17.11 & 19.91 & 23.76 & 22.31 & 23.89 \\  
    \textbf{\checkmark} & $-$ & 72.91 & 48.22 & 35.44 & 51.98 & 40.16 & 49.74 \\
    $-$ & \checkmark & 37.07 & 19.72 & 21.58 & 27.33 & 23.48 & 25.04 \\
    \checkmark & \checkmark & \textbf{75.18} & \textbf{50.06} & \textbf{39.17} & \textbf{54.59} & \textbf{43.97} & \textbf{52.39} \\
    \bottomrule
  \end{tabular}
  }
  \vspace{-3mm}
\end{table}


\noindent\textbf{The effects of Fine-tuning CLAP.}
While CLAP excels in speech transcription and retrieval, it lacks mechanisms to model emotional attributes, limiting its effectiveness in SER. To address this, we fine-tune CLAP on large-scale emotional speech corpora (Tab.~\ref{tab:datasets}) to establish a stronger prior for emotion recognition. As shown in Tab.~\ref{tab:ablation_finetune}, CLAP (Laion-AI) achieves only 23.89\% accuracy on SER, highlighting its inability to capture emotional cues without fine-tuning. Fine-tuning improves performance by 25.86\%, demonstrating its effectiveness in enhancing CLAP’s alignment of emotional information across modalities. This step establishes CLEP-DG as a stronger foundation for SER, improving generalization across emotional contexts.


\noindent\textbf{The effects of Acoustic Context Prompt Tuning.}
As shown in Tab.~\ref{tab:ablation_finetune}, applying ACPT alone improves performance over the CLAP model, demonstrating that augmenting textual representations with acoustic prompts enhances emotion alignment. When combined with fine-tuning, ACPT further improves CLEP-DG’s accuracy by 28.5\%, confirming that exposing the model to diverse acoustic contexts during training enhances its robustness to unseen environments. This result highlights the effectiveness of text-driven augmentation in improving generalization across varied recording conditions.

\begin{figure}[!t]
  \centering
  \vspace{-7mm}
  \includegraphics[scale=0.32]{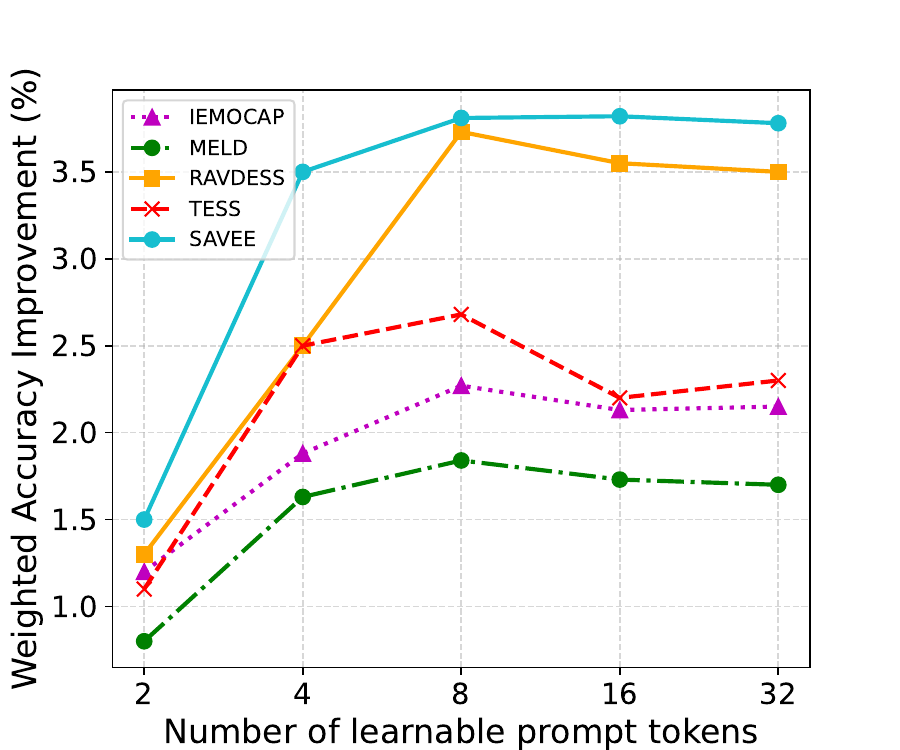}
  \caption{ Performance improvement with different numbers of prompt tokens compared to the fine-tuned baseline.
  }
  \label{fig:tokens}
  \vspace{-4mm}
\end{figure}

\noindent\textbf{The length of the learnable prompt.}
Fig.~\ref{fig:tokens} shows the effect of prompt length on weighted accuracy. We evaluate 2, 4, 8, 16, and 32 learnable tokens, observing that larger token sizes improve performance. Accuracy rises sharply up to 8 tokens, then plateaus. While longer prompts (16, 32) sustain high accuracy, they yield diminishing returns. Thus, we select 8 tokens as the optimal length, balancing performance and efficiency.

\begin{table}[htpb]
  \centering
  \caption{Ablation study on the classification loss $L_{\text{class}}$ used for training a linear classifier in the proposed framework.}
  \vspace{-2mm}
  \label{tab:arc_loss}
  \resizebox{\columnwidth}{!}{
  \begin{tabular}{lcccccc}
    \toprule
    $L_{\text{class}}$ & IEMOCAP & MELD & RAVDESS & TESS & SAVEE & Avg. \\
    \midrule
    Softmax  & 74.59 & 49.05 & 38.37 & 53.93 & 42.58  & 51.54 \\
    \textbf{ArcFace} & \textbf{75.18} & \textbf{50.06} & \textbf{39.17} & \textbf{54.59} & \textbf{43.97} & \textbf{52.59} \\
    \bottomrule
  \end{tabular}
  \vspace{-4mm}
  }
\end{table}
\noindent\textbf{The effects on the classification loss.} Tab.~\ref{tab:arc_loss} compares the impact of Softmax and ArcFace on training the linear classifier. ArcFace consistently improves accuracy across all datasets, achieving an average gain of 1.05\%. These results demonstrate that the angular margin constraints in ArcFace enhance class separability, resulting in a more discriminative feature space and a stronger decision boundary for speech emotion recognition.

\section{Conclusion}
We present CLEP-DG, a CLAP-based framework that enhances text-to-emotion audio alignment for SER. Integrating Acoustic Context Prompt Tuning (ACPT), CLEP-DG improves generalization across diverse acoustic environments without additional labeled speech data. Leveraging cross-modal transferability, it bridges textual supervision and audio-based inference, reducing data dependency while improving robustness. Experiments across multiple benchmarks show that CLEP-DG outperforms existing CLAP-based approaches, achieving state-of-the-art accuracy and generalization.

\newpage
\newpage
\bibliographystyle{IEEEtran}
\bibliography{main}

\end{document}